# Will molecules ever sit at the thermoelectric table?

*8 technological milestones.*

Andrea Gemma and Bernd Gotsmann
IBM Research – Zurich

26 August 2020

## Abstract

Molecular junctions comprise small molecules on the order of one or a few nanometers in length, chemically attached to two electrodes of metal or semiconductors. They have recently been identified as promising candidates for materials/devices for thermoelectric energy conversion. Molecules have the potential to act as sharp energy filters for electrical currents and could outperform other materials considered for thermoelectric energy conversion so far, thereby providing a wide-spread technological relevance thermoelectrics could not achieve to date. However, there is a clear gap between predictions and demonstrations in the literature studying thermoelectric molecular junctions. This opinionated review seeks to highlight necessary steps needed for this approach to transition from fundamental research into a viable technology. Some important conjectures, on which the study of molecular junctions is based, are reviewed.

## Introduction

A molecular junction is a device, in which one or several molecules are contacted individually by conducting electrodes. Typically, one employs small organic molecules with delocalized pi-electron system along molecular backbone, into which and out of which charge carriers tunnel to and from metallic electrodes. The electrodes also act as thermal reservoirs, see Fig. 1.

The most interesting transport regime of molecular junctions comprises coherent charge transport, which typically happens in relatively small molecules of maximum a few nanometers in length. In such systems, transport is governed by a single, energy-dependent transmission coefficient of charge carriers. A transmission of $\tau = 1$ corresponds to a fully open conductance channel of one conductance quantum 1 $G_0$. In longer molecules hopping transport predominates, which follows different physics, and, importantly, has worse charge conductance properties. A typical candidate molecule attaches to gold electrodes with gold-thiol bonds. However, other electrode materials and coupling mechanisms have been studied, for example pi-pi stacking onto graphene electrodes or silanization of silicon (oxide) surfaces.

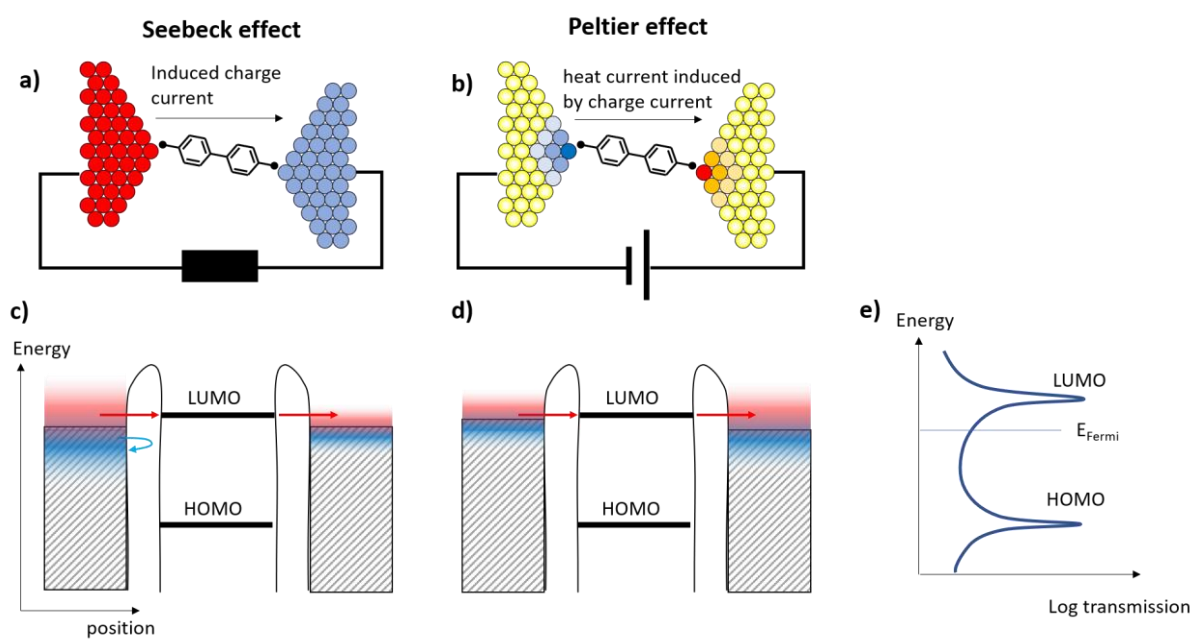

**Figure 1: Schematic of a molecular junction acting as energy harvester using the Seebeck effect (a) or as cooling circuit using the Peltier effect (b) energy diagram. The respective energy-dependent transport of charge carriers from metal electrodes via tunneling barriers and molecular orbitals is depicted in (c) and (d) for the two effects, respectively. The energy-dependent transmission function needs a high transmission for high electrical conductance and a large energy dependence of transmission at the Fermi energy for large Seebeck coefficient. Unfortunately, molecular junctions tend to have their Fermi energy rather mid-gap between the highest occupied molecular orbital (HOMO) and the lowest unoccupied molecular orbital (LUMO), resulting in small Seebeck and small conductance.**

After the first claimed single-molecule transport measurements in the 1990s [1], the field rapidly grew, focusing initially on logic applications [2,3]. Today, molecules are only rarely mentioned any more as replacement technology for highly-scaled semiconductor electronics. New research directions have emerged, and energy conversion appears to be particularly popular. The study of thermoelectric effects in molecular junctions has been going on for about a decade following both initial experimental measurements of Seebeck coefficients (also called thermopower) and theoretical description [4,5,6]. Now several tens of research groups are active in the research area. It is clearly a well-functioning citation network, and from the high citation levels alone, one cannot conclude yet on technological relevance.

It is important to not confuse molecular junctions with conventional organic electronics. Organic matter for electronic applications has been a field of active research for decades. It has found applications such as plastic electronics for product tags or organic solar cells, as well as a wide market penetration in organic light emitting diode displays. However, molecular electronics and organic electronic share important advantages and disadvantages over inorganic materials. The advantages include the comparatively low price, relatively easy mass production and the high flexibility in terms of form factor. Well-known disadvantages are the low electrical conductivity and low thermal stability.

There are important differences also. In organic electronics, we are mostly talking about thin films on the order of microns of amorphous organic materials. Molecular junctions, however, comprise small molecules on the order of one or a few nanometers in length, chemically attached to two electrodes of metal or semiconductors. Therefore, molecular junctions are more difficult to handle. The integration into thin-film assemblies for most practical technological applications is not achieved to date. On the other hand, there are key advantages of molecular junctions. The charge transport can be precisely tuned, and various quantum-mechanical transport effects are directly accessible.

In the following, we sketch some of the most important milestones on the path to bring molecular junctions to technological readiness within the context of thermoelectric energy conversion. The first section describes the fundamental science aspects that lead us to believe that research is well invested into molecular junctions for this purpose. The second section will consider the integration aspects of molecular junctions into devices and the thin-film problem. Finally, some boundary conditions for technological applications are given.

## Physics aspects: Why should molecular junctions be interesting for energy conversion?

Several recent and ongoing research projects receiving significant government funding are based on the notion that molecular junctions are potentially good energy converters. These include the EU H2020 projects EFINED, QuIET and MOLESCO, the ERC projects CHEMHEAT, the UK-based projects QuIET and QSAMs, and numerous personal grants in the US, China and Europe. Many of them are motivated by predictions of a high thermoelectric conversion efficiency near equilibrium, as expressed using the figure of merit $ZT$ through thermal conductance $k$, electrical conductance $G$ and Seebeck coefficient $S$ of a single molecule junction. A value of $ZT = G S^2 T / k$ on the order of 1 can be achieved with solid materials such as $BiTe_x$, SnSe or Heusler alloys, at room temperature, with the highest laboratory reported values around $ZT = 3$ [7,8,9]. The hope is that molecular junctions can reach similar or higher values of $ZT$. How $ZT$ translates into an energy conversion efficiency is shown in Fig. 2, in which the cold reservoir is set to room temperature.

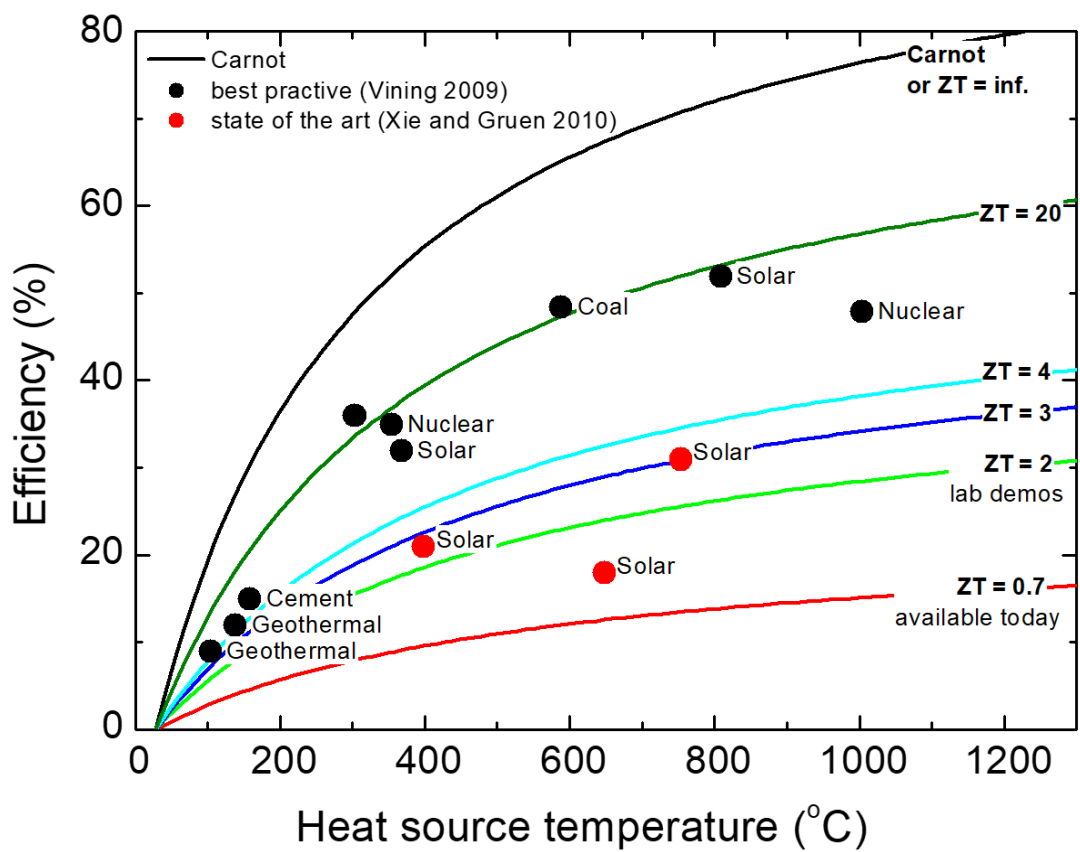

**Figure 2: Energy conversion efficiency versus temperature for different figures of merit.** Adapted from [10] and [11]. Notably, existing power plants are significantly below the optimum Carnot efficiency. This is in part due to the fact that the power is another important criterion. At the Carnot limit, a machine has no conversion power. For simplicity we discuss here *ZT* in terms of maximum efficiency and refer to the literature for a further discussion[12].

### Seebeck coefficient

The promise of a high Seebeck coefficient, also called thermopower, is the main motivation to use molecular junctions for energy conversion[13,14,15]. The Seebeck coefficient is the ratio of open circuit voltage measured over a sample divided by the temperature difference between the electrodes, and typical values are in the range of single to hundreds of µV/K, depending on material. The fact that the electronic states in molecules are quantized and transport can occur resonantly, i.e. involving HOMO or LUMO of a molecule (i.e. the highest occupied and lowest unoccupied molecular orbitals, respectively), means that there can be strong energy-dependence of transport. This is a prerequisite of the thermoelectric effect[16].

However, typical molecular junctions of a single molecule chemically bound to metal electrodes tend to have a Fermi energy placed well in the gap between HOMO and LUMO[17,18]. Consequently, both electrical conductance and Seebeck coefficient are typically rather low. Most measurements of molecular junctions have yielded Seebeck coefficients of only few µV/K [19,20]. Relatively high values of tens of µV/K have been reported for $C_{60}$ and other fullerene junctions[19].

There are two interesting approaches to deal with the problem of low Seebeck coefficients. One comprises efforts to move the Fermi energy towards either HOMO or LUMO levels. This is difficult and some of the proposed approaches involve molecules comprising charged or radical groups or (electro)chemical gating [21,22,23,24]. A second approach, which has gained considerable attention recently, is the design of molecular systems such as quantum interference effects. If there are several paths for charge carriers travelling along the delocalized pi-electron systems, then destructive interference of their wave functions may occur. As a consequence, the transmission coefficient as a function of energy can have very sharp features between the HOMO and LUMO levels[25,26,27], as shown in Figure 3. First experimental studies report evidence that quantum interference does exist in single molecules [28,29]. It is still to be demonstrated that it can lead to enhanced Seebeck coefficients, with values exceeding 100 µV/K at reasonable conductance.

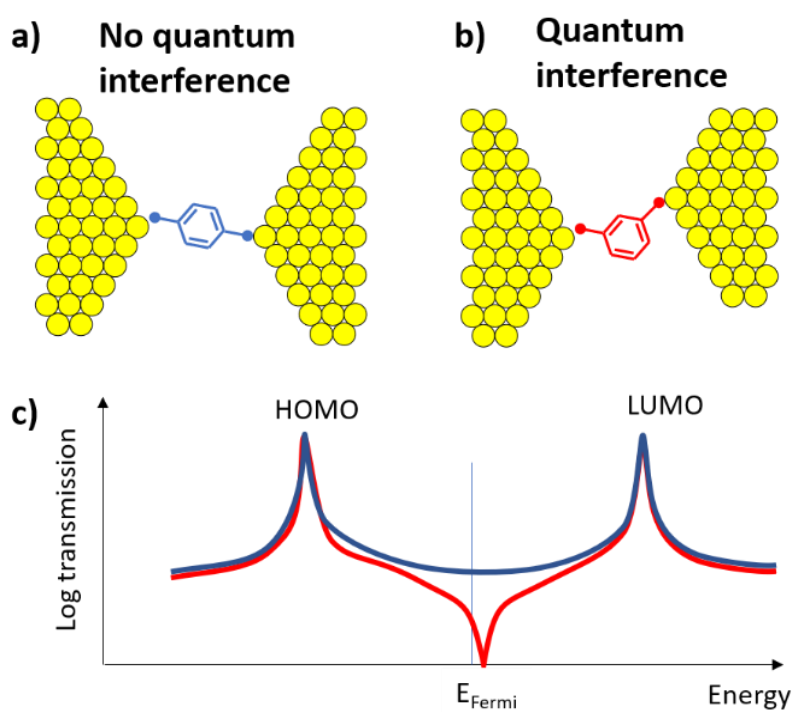

**Figure 3. (a) Schematic of molecular junctions without (a) and with (b) molecular interference. The corresponding transmission spectra do or do not show a feature in the HOMO-LUMO gap, which could be exploited for an enhanced Seebeck coefficient. In this schematic example, the red molecule junction would have a larger Seebeck coefficient (through the slope of transmission versus energy, τ(E)) and a lower conductance (value of τ), with respect to the blue molecule junction.**

### Electrical conductance

The rather poor electrical conductance of individual molecular junctions is and has been regarded as one of the major reasons against a technological feasibility of molecular electronics. Individual molecules, designed for high conductance, may reach conductance values of $10^{-2}$ $G_0$. However, these are typically short molecules (on the order of 1 nm) with relatively little freedom to tune the transmission properties. More typically, however, molecular electronics experiments at the single molecule level find rather $10^{-3}$ to $10^{-5}$ $G_0$, for molecules of a few nm in length. Even longer molecules tend to have incoherent transport properties and are therefore less useful for energy filtering approaches.

> **Milestone 1:** Demonstrate a molecule junction with a power factor of $GS^2 > 10^{-14}$ W/K$^2$.

It is clear that both Seebeck coefficient and electrical conductance play a role in thermoelectric applications. In poor conductors, large Seebeck coefficients can be observed. However, these are of little relevance. Therefore, it is useful to consider the power factor $GS^2$. (Note that in bulk materials one oftentimes considers $\sigma S^2$ instead, using the electrical conductivity $\sigma$.). In all the applications, like cooling, where the maximum power is more important than efficiency, the power factor is an equally relevant quantity to $ZT$ [12,30]. Milestone 1, reaching $10^{-13}$ W/K$^2$, translates, for example, to a Seebeck coefficient of 360 µV/K at low bias for a conductance of $10^{-3}$ $G_0$. To compare with Bismuth Telluride bulk materials, which have a power factor $\sigma S^2 > 30$ mW/(K$^2$m), we consider a molecular length of 3 nm and an area of 5x10$^{-19}$ m$^2$, and reach $GS^2 = 6\times10^{-13}$ W/K$^2$. Milestone 1, although ambitious, does not yet compete with existing technology for cooling power, but may be competitive for efficiency, see below.

For large bias applied to a cooling system, reaching non-equilibrium conditions and leaving the linear response regime, however, the Seebeck coefficient and therefore the power factor are not well-defined. Then an overall cooling power per molecule is a more appropriate measure. Here a goal is to reach a cooling power sufficient to compete with thermoelectrics, which today reaches a temperature difference on the order of 60 K using a single thermoelectric layer.

Thermal conductance

For molecular systems studied in the literature so far it appears that the phonon-based thermal conductance dominates thermal conductance. The thermal conductance of a molecular junction, $K$, with a single molecule is typically on the order of a few 10s of pW/K [31,32], the electrical thermal conductance, in contrast, is expected to contribute sub-pW/K. This is rather problematic in view of reaching a high $ZT$, especially when comparing to good bulk thermoelectric materials in which the dominance of phonon thermal conduction can be avoided or reduced.

> **Milestone 2:** Achieve phonon thermal conductance below 10 pW/K for a molecule with electrical conductance of at least $10^{-3}$ $G_0$. In the long run, both numbers should be decreased and increased by one order of magnitude, respectively.

There are several properties of phonon conduction that invite research into a reduction of thermal conductivity. First, the molecular structure may comprise side-groups that are electrically inactive but may reduce phonon conduction [33,34]. Secondly, intra-molecular interference effects may be exploited [35, 36]. Thirdly, a variation of vibration frequencies along the molecular backbone may be achieved for lower phonon transmission [37]. For example, one may exploit the use of heavy atoms in combination with light carbon backbone. Finally, phonon mismatch to the metal electrodes can be optimized. Since the phonon thermal conduction is mediated with an entire spectrum of phonons distributed using the Bose-Einstein distribution, it may seem that there is less handle to engineer phonon conductance as compared to charge transport. However, from the list of proposed mechanisms to reduce phonon conduction, it appears we are only starting to see the potential of phonon tuning of molecular junctions.

## Efficiency

To be able to study systematically the efficiency of candidate molecules for thermoelectric applications, we need to be able to measure the relevant properties $S$, $G$ and $K$, and thereby $ZT$ and the power factor. So far, there is only one molecule that has been experimentally measured in a molecular junction on the single-molecule level by two independent groups[38,39]. There is clearly need for instrument development to accelerate such studies. Although, consensus is gradually formed for many specific molecules studied in different teams, we must not forget that differences between independent results are easily half an order of magnitude in electrical conductance. The fact that the experimentalist is blind to the exact configuration of the molecules in the junction and the freedom to interpret statistical analysis results is still a major limitation in this research field. With interest we are looking forward to more results using machine learning approaches to classify statistical data.

> **Milestone 3:** a) Demonstrate the ability to measure and predict all relevant thermoelectric properties (electrical, thermal and thermoelectric transport) of a specific molecular junction in a reproducible manner at the desired temperature.
> b) Demonstrate a figure of merit $ZT > 3$.

Milestones 1 and 2 combined would correspond to a $ZT$ of 0.3, which falls short of the state of the art. To estimate an attainable $ZT$ value for a material class, it is a dangerous and oftentimes misleading custom to combine individual best values for $S$, $G$ and $K$. However, not only G and S are highly interdependent, also K and S are. Maybe due to the lack of experiments and predictions of K, it has become a custom in the molecular electronics community literature to report an "electronic ZT", or $ZT_e$, rather than the true $ZT$. For $ZT_e$, only the thermal conductance through charge carriers is considered and the phonon contribution neglected. In this way very large efficiencies can be implied, which can be rather misleading. Furthermore, the validity of the Wiedemann Franz relationship is often assumed between electrical and thermal conductance, which may not be valid in all molecular junctions, even if phonon transport would be neglected. In this context it is interesting to note that assuming the Wiedemann Franz law ($K = L_0 T G$), the efficiency only depends on the Seebeck coefficient, $ZT = S^2/L_0$, which would reach 0.4 for $S = 100$ μV/K. Here $L_0$ is the Sommerfeld value.

It is safe to conclude, that there is a huge discrepancy between predictions of molecular systems in terms of their thermoelectric efficiency and what has actually been demonstrated. But where shall we look for improvement? Let's assume phonon thermal conductance could be suppressed to 1 pW/K. To have charge carriers dominating K, we estimate a conductance of $G > 10^{-3}$ $G_0$. To reach $ZT = 3$ (milestone 3) we then require S = 270 μV/K, i.e. a power factor of $10^{-14}$. Whether a $ZT$ of 3 is sufficient to motivate technological application remains doubtful, as we will discuss in the following, but would at least motivate further consideration.

## Materials science aspects: The fabrication of molecular films maintaining the advantages of molecular junctions

A device based on a single molecule junction for energy conversion processes is currently not realistic in technology and may not be needed in the foreseeable future. However, a device using an ensemble of molecular junctions in an array to allow parallel operation and, more importantly, integration into a solid-state device is clearly envisioned.

From regarding single molecule junctions, however, it is not obvious that films of molecular junctions will maintain the advantageous properties of single-molecule junctions. The conformational and vibrational freedom of molecules can be influenced by neighboring molecules and take influence on both phonon and charge transport [40]. Furthermore, coherent transport effects may occur, in which, for instance, a phonon from the bulk electrodes uses two molecules to transverse.

> **Milestone 4**: Measure charge and thermoelectric transport across arrays of molecular junctions, maintaining or improving thermoelectric properties measured in single molecule junctions.

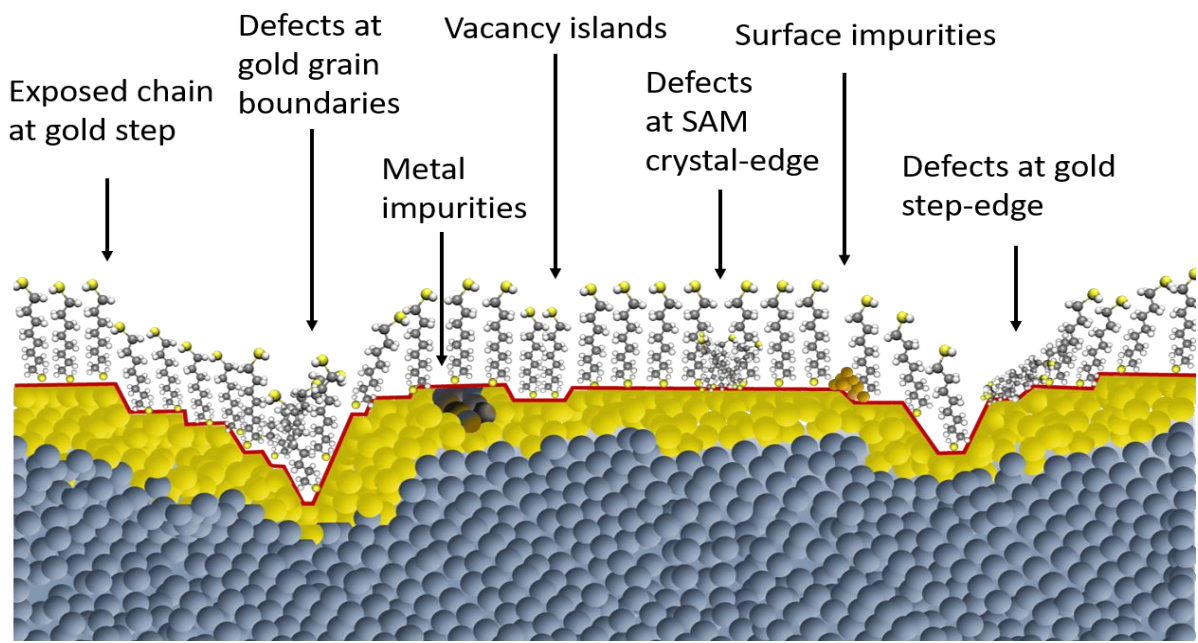

**Figure 4:** The perfect crystalline arrangement of self-assembled molecular monolayers on surfaces can have a number of different defects that may impact technological exploitation. Shown here are some prominent examples as observed in the well-studied alkane SAMs on gold surfaces in a schematic representation. Adapted from [41].

The measurement of molecular carpets may not suffer from having to measure tiny thermal conductance signal on the order of pW/K. However, when measuring molecular ensembles using scanning thermal microscopy or thermoreflectance, large systematic errors have been reported. In SThM, there is uncertainty of the exact contact area and series thermal resistances in tip and sample. Thermoreflectance measurements need a top-contacting electrode on the molecules, and typically it has to be assumed that only a fraction of ca. 50% of the molecules actually binds[42]. The reasons for this are the multitude of different defects in molecular films[41,43], as shown in Figure 4, which will likely diminish device performance.

It is important to resolve such issues and to engineer thermal transport in molecular films (milestone 4). The effect of molecular surrounding or groups of molecules on the transport may well be an opportunity to tune transport in improve energy conversion properties. An obvious goal may be to reduce phonon thermal conductance. As phonon transport along short molecules has been shown to be coherent, interference effects may be a first point of examination. A scientific study on molecular pairs may be a way of comparing experimental to simulation data. Experiments on molecular films, namely self-assembled monolayers sandwiched between electrodes will be an important step.

> **Milestone 5**: Thermal quantification and engineering of phonon conductance of molecular films. Quantify the role of radiative thermal transport in the presence of organic molecules for nanometric electrode gaps.

Finally, the role of near-field electromagnetic radiation in heat transfer between closely separated electrodes needs to be addressed. Currently, the so-called extreme near-field across gaps of only few nanometers is highly debated in the community [44,45]. Experimental data suggests that the presence of molecular systems may have significant influence[44] and could reach similar orders of magnitude as the phonon conduction across the molecules. It is therefore of paramount importance to address this issue experimentally and theoretically (milestone 5).

> **Milestone 6:** Demonstrate an attractive thermoelectric efficiency (e.g. $ZT > 3$, for the harvesting application) for thin film / monolayer molecular junction.

Only when these issues, mainly experimentally, have been addressed, we can reach functional demonstrations of molecular carpets in view of technological application. The determined efficiency (Milestone 6) will decide whether it makes sense to start a development and testing phase to continue work on energy harvesting. Cooling applications may again be more forgiving with respect to requirements.

In quest for improved efficiency, it is worth mentioning the existing efforts on replacing the rather resistive coupling of the molecules via the gold-thiol bond. Quite popular appears to be the use of graphene electrodes with pi-stacked anchor groups. The planar geometry requiring a carrying substrate, however, is not suited for harvesting applications and limits cooling application scenarios. More relevant approaches include using other metals and alternative anchor groups. Future work will comprise fabrication issues etc. In terms of science, however, the next step after a successful thermoelectric demonstration on a single film will be to measure and understand endurance issues. A film will need to be temperature-cycled many times during an anticipated ten-year lifetime. Numerous methods for predictive testing of materials for life-time analysis are ready to be applied to molecular junctions.

> **Milestone 7:** a) Determine the endurance and cyclability of a molecular thermoelectric material in view of a 10-year operation at cycling temperatures.
> b) Create and fit a physics-based model predicting the degradation mechanism of molecular junctions.

### Engineering science aspect: The integration of molecular films into working devices without destroying the efficiency

Having discussed the development of functional prototypes above, we now turn to engineering aspects of molecular films. Some of these aspects can be studied already resulting in valuable feedback to fundamental research of molecular junctions in terms of potential and actual technological need of a solution. The most pressing issue foreseen by the authors is sometimes called "thin-film issue" of thermoelectrics and will be described next.

One must be careful choosing or discarding molecular systems for energy conversion based on single-molecule data alone. The most likely implementations use an assembly of molecules, such as a self-assembled monolayer, as a functional unit. In such layers, for example, a $10^{-4} G_0$ value at a packing density of 0.21 nm$^{-2}$ translates into ~$10^9$ $S$/m$^2$. While this is an attractively large value, we have to remember that the relevant value for efficiency is the ratio between electrical and thermal conductance of the junction (entering $ZT$), and the thin-film problem, as discussed above. A thermal conductance of 30 pW/K per molecule would translate to ~$10^7$ W/K/m$^2$. This is a typical order of magnitude for (good) thermal interfaces in general, which is alarming for the following reasons.

There are challenges associated with integrating thermoelectric materials into a working device for energy harvesting or cooling. Roughly, half of the efficiency expected from the $ZT$ value of a material is lost in the functional device even using bulk materials. There are several reasons for this. First, for electrical impedance matching, one uses a series of miniature Seebeck (or Peltier) elements operated electrically in series and thermally in parallel. Both p- and n-type materials are used requiring a certain complexity of contacts and wiring. Secondly, the thermal contacts between thermo-electrically active materials and the thermal reservoirs induce a drop across the contacts. Thereby valuable temperature difference is lost for energy conversion. In conventional thermoelectric elements, this temperature drop is to a certain extend manageable through optimizing the aspect ratio of dimensions of the thermoelectric elements. However, for thin film thermoelectrics, like the ones that a molecular monolayer junction would be, the situation is more severe. Series thermal resistances carry more relative weight in thin-film thermoelectrics.

We can use a rough estimation to explain the severity of the effect. The thermal resistance $R_{molec}$ per unit area of a densely packed molecular layer as estimated above will be on the order of ~$10^{-7}$ Km$^2$/W. This is to be compared to the other thermal resistances involved. These are due to the electrical contacts to the molecules $R_{contact}$, and electrical isolation layer $R_{isolation}$ and the thermal interface $R_{interface}$ to the thermal reservoirs. A metal contact of 100 nm thickness has a negligible value of about $R_{contact}$ = 5x10$^{-10}$ Km$^2$/W. However, a 100 nm silicon oxide insulator already has $R_{isolation}$ = 7x10$^{-8}$ Km$^2$/W, and we need to consider at least four thermal interfaces (between electrical contacts and insulator and between insulator and electrical reservoirs), all of which are at best on the order of ~$10^{-8}$ to $10^{-7}$ Km$^2$/W. Consequently, the temperature drop across the molecular layer will therefore be only few % to a few 10s % compared to the one available in the thermal reservoirs.

> **Milestone 8:** Demonstrate basic concepts of thermal impedance matching of molecular systems

An engineering effort addressing this thermal impedance issue needs to be made (milestone 8). An obvious handle to turn would be the packing density of the molecular film. This would require exploring ways to stabilize two closely separated electrodes, something which should be possible. A second important approach is to use literally or effectively a multilayer of molecular junction carpets. A beautiful approach is the use of nanoparticle assembly[46,47]. Research needs to be made trading off the losses of efficiency through non-optimal binding of all involved molecules and the increase of thermal resistance of the film per unit area. Such arrangements will have specific additional demands on mechanical stress during thermal cycling.

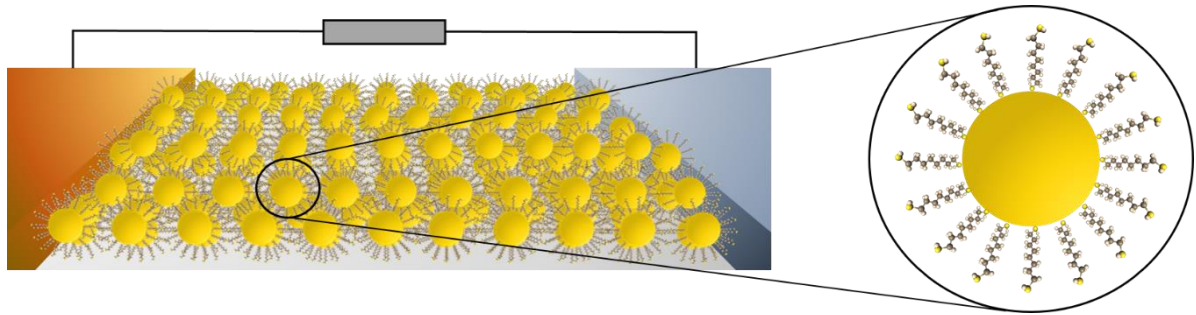

**Figure 5: Nanoparticle array as using in research of molecular electronics. Such arrangements could provide a way of scaling up molecular junctions to larger dimensions and ultimately avoiding the thin-film problem.**

## Technological use cases: It is not only about the figure of merit ZT

It is out of question that a high conversion efficiency is always an advantage for the use of thermoelectric materials in all applications, and that the power factor is an important parameter for cooling applications. However, scientific literature oftentimes ignores that there are a number of parameters governing a technological product. These include the toxicity of the materials, the cost of fabrication, the lifetime, the robustness (against temperature variation, radiation, vibration, …), the weight of the module, and many more. We will turn to a list of applications that have been discussed in the context of thermoelectric energy conversion, mainly for thermoelectric electricity generation (TEG). A decisive factor is always the availability of other competing technological solutions. Higher cost can sometimes be traded against better technological performance.

### Mass energy production from primary heat sources

The potential of thermoelectrics as replacement technology of gas turbines in large power plants or in solar energy (either photovoltaics or solar thermal) have been controversially discussed already [10,49,50]. In any case, the existing technology is well-established and, so far, thermoelectrics have not been implemented. In addition to the lack of demonstrated efficiency, organic or molecular thermoelectrics will not easily stand the high temperatures of the hot reservoir requesting. Therefore, molecular junctions fail due to low **thermal stability** and **conversion efficiency**. On an interesting side-remark, even gas turbines do not endure the hot-side temperatures of nuclear power plants, and the steam needs to be cooled before entering the turbine. Thermoelectric materials that can deal with temperatures above 1000 C would certainly be interesting. A second side-remark, to work with large temperature differences, thermoelectric converters can be cascaded, such that a certain temperature range is seen by a particular material. In such a scenario, organics could play the role of a near-room temperature component of such a stack.

### Waste heat

Waste-heat recovery concerns heat sources of moderate temperatures of tens to hundreds of degrees and, depending on the heat source, moderate power. Typical heat sources are industrial processes or the low-temperature waste heat of large power plants. In contrast, energy conversion from very small heat sources are sometimes referred to as energy scavenging. For very large heat sources in terms of power (not temperature), existing solutions are mechanical turbines and dynamos. A natural focus for thermoelectrics is therefore relatively small heat sources, for which no economical energy harvesting solution exists.

For an industrial installation a relevant metric is the installation **cost per Watt** of energy conversion at an existing industrial plant. An interesting recent study commissioned by the Swiss Federal Office of Energy [51] considers a target of about 3 Euro per Watt for an installation, basing the estimates on a market price of 0.15 Euro/kWh and a payback time around 5 years. The study concludes that there is a factor of 5 missing to make thermoelectric installations profitable. Another independent study [52] arrives at a similar conclusion, that the total cost for the installation of thermoelectric converters must be decreased at least by 7 times to make it appealing for the industrial settings. Having said that, the political landscape may change in several countries making such installations profitable sooner. In terms of unit costs, a thermoelectric converter comprises expensive materials, usually $BiTe_x$, and expensive assembly costs. Therefore, by trading off materials and assembly costs versus efficiency, there may be a chance even for current thermoelectric materials. For example, silicides are estimated to cost only 4 US$ per kg in comparison to BiTe, which is on the order of 100 USD$/kg [53]. A molecular system may have an advantage through the small amount of material needed but needs to demonstrate the module fabrication cost and lifetime.

### Automotive

Automotive waste heat recovery is a popular special case of waste heat recovery. An engine burning fossil fuel has a large exhaust temperature and a comfortable 20 to 200 kW engine power. In a car the usable electricity for car electronics can be on the order of 600 W. A relevant metric here, however, is the **Watt per kilogram** of installation. The additional power needed to accelerate the additional weight of a thermoelectric converter can be estimated to be about 60 W per kg. A second important aspect is that the heat needs to be removed relatively quickly with the exhaust gas from the engine. This necessitates a heat exchanger of significant weight. If we assume 10 kg for a heat exchanger, then the first 600 W of conversion power is lost to the engine. While there may be use in larger cars, we note that mechanical engines may also be used to recover waste heat from automobiles. More importantly, the industry has a focus in developing electric cars, in which not sufficient waste heat will be produced. It appears this application is not particularly promising, and molecular systems may not be applicable for reasons of the high temperatures involved.

### Mobile devices

Another popular vision for thermoelectric conversion is the direct use of body heat to charge mobile devices such as sensors or even smartphones. To give a reference, the stand-by power of a smartphone is on the order of 70 mW, which is an interesting target [54].

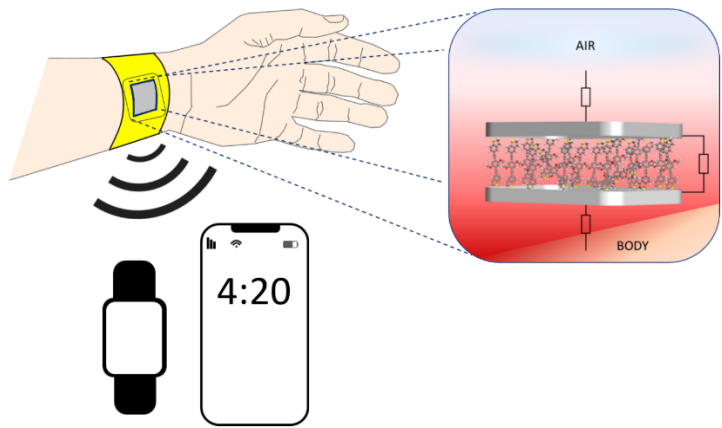

**Figure 6: Harvesting energy for IoT from body-heat using a thermoelectric module. The thermal resistance associated with the interface with the surrounding air can lead to very large temperature drops, which are not exploited using the thermoelectric energy conversion.**

A human body dissipates roughly 50 mW per 10 cm² of the body surface, a value certainly worth considering as an energy source (up to a point the person feels cold). With an efficiency of 1.6 % ($ZT$ = 3 and $\Delta T$ = 15 K) this would translate into about 1 mW of power. However, this assumes that the full temperature difference between body and ambient of about 15 K can be used. It looks much less promising, if one considers the cascade of thermal resistances in series, as discussed above. In this particular application, the situation is even worse and cannot easily be improved with the measured discussed above. The problem is the thermal interface to the cold reservoir i.e. the ambient air, which couples by convection about 20 W/K/m². Without dedicated and bulky cooling structures[55], there will be a more realistic temperature drop over the thermoelectric around 0.1-1 K , reducing the delivered power to a maximum of 5-50$\mu W$ for a 10 cm² patch. Considering that a $ZT$ = 3 is hard to achieve outside the laboratory, even for a basic optimized design, results are not better, with a temperature drop of ~5 K and an output power of ~1$\mu W$ [56] .For comparison, photovoltaics harvesters for in-door lighting reach on average 5 $\mu W/cm^2$ [57]. It appears, the energy harvesting application from body-heat is built on an unrealistic dream.

### Sensing nodes for the IoT

In the next years, new interesting possible domain of application for thermoelectric materials used as generators could arise from the field of distributed, intercommunicating sensing networks (the so-called internet of things, IoT). In this framework a variety of interrelated devices such as computers, sensors or in more general terms "nodes", will have the ability to receive and transmit data autonomously, over a network, without human-to-human or human-to-device interaction[58,59,60]

In this respect the main advantage of thermoelectric technology consists on its high reliability and endurance, no use of chemicals and/or moving parts, with a lifetime that often overcomes the one of the devices to be powered. On the other hand, the major drawback, as already discussed in the previous paragraphs, is represented by the poor conversion efficiency and thermal coupling, yielding a limited power output. However, for small sensing nodes the amount of converted power could be adequate to replace or at least flank ordinary storage solutions (batteries) or complement different types of renewable-energy supplies (e.g. photovoltaic).

In Table 1 some examples of power supplies for different commercially available sensing nodes (SNs) are given. It is important to notice that the largest part of the power is actually consumed by the transceiver module (~mW)[61], and not by the microprocessor. In fact, the extraordinary progress made by the IC industry has led today to the availability of ultra-low power microcontrollers with a power demand down to few tens of $\mu$W in their active state[61]. Additionally, the fact that microprocessors spend most of their time in sleep-mode makes their power consumption negligible compared to the one of the transceivers. All that makes difficult to estimate the power needed by a SN, since, even for a given architecture, it strongly depends on the amount and rate of data transmitted.

| Device | Range (m) | Components | Power Supply | Device area (cm²) | Application |
|---|---|---|---|---|---|
| **WeC** | 4.6 | MCU, Storage, Light and Temperature sensors | Coin cell, 24mW | 6.25 | Environmental sensing |
| **Rene2000** | 30.5 | MCU, Storage, GPIO, SI | Battery, 24mW | Not available | Environmental analyses |
| **Mica** | 61 | MCU, Storage, GPIO, SPI | 2xAA, 27mW | 17.36 | Environmental sensing, |
| **Mica2Dot** | 152 | MCU, Storage, GPIO, SPI | 2xAA, 44mW | 7.75 | Environmental sensing |
| **Imote2** | 91.5 | MCU, Storage, Camera, SI | 3xAA, 86.8mW | 16.92 | Environmental sensing |

**Table 1: Power supply and main characteristics of commercial sensing nodes. Adapted from** [62]

The actual output power that a microTEG outside laboratories depends strongly on the varying ambient conditions. The output power therefore not only depends on the efficiency $\eta$ and *ZT*, but also on the coupling with the environment, i.e. the available thermal budget and footprint/contact area. The fact that, heat fluxes and temperature differences might considerably fluctuate in typical context of use, prevents the use of TEGs to power sensors, which require a steady and reliable operation. In this respect Lithium ion batteries are still a safer option for small nodes. Most of the times, additional designing steps are required and consequent circuitry solutions must be adapted to the specific deployment conditions of thermoelectrics.[63]

To give some examples, for embedded road sensors, temperature differences between asphalt and road subgrade soil are usually between 10-20 K[64,65], and it has been experimentally shown that only with an engineered design and proper footprint the temperature gradient could be enough to sustain power output of ~$10^{-2}$W [64,65,66], allowing the operation of SNs in Table 1. A similar scenario could be represented by the case of small bipolar thermal gradients occurring at the ground-to-air boundary. Even in this case, through an optimized design, the self-sustainable operation of an environment sensing application with a 550 µW power demand has been demonstrated[67]

Albeit from a theoretical point-of-view TEGs would be a convenient solution to power SNs for many applications, some more practical reason can explain the inertia of the industry to exploit and further develop thermoelectric solutions for the IoT. As previously mentioned, in case of a thermoelectric power supply, the overall SN design must be re-optimized in order to improve the heat uptake at the hot side and the heat dispersion at the cold side, considerably increasing production time and final price. Also, the overall price of a SN network is, in the first place, fixed by the capital cost (electronics, installation…), thus for big the impact of TEGs can be from an energetic point-of-view, it might turn to be negligible on the cost as a whole[62]. However, it needs to be studied whether TEGs could significantly reduce service costs for replacing batteries. So far, it seems that thermoelectric generators only represent a convenient solution in case of forbidding locations, or when parts substitution/repair plays a major role.

## Chip or device cooling

In general, mechanical engines are also the competitor for cooling applications like refrigeration in households for food or air conditioning. With similar arguments as above, there appears little motivation to propose thermoelectrics as an alternative except for small form-factor niche applications.

However, stabilizing the temperature of optical components is one of the most important uses of thermoelectrics today. Here cooling power, and thereby the **power factor** is a relevant metric, complementing conversion efficiency. Motivated by the nanoscale nature of materials studied today for thermoelectrics, it appears interesting to consider cooling even for local cooling for microscale electrical devices. An issue with this approach is surely the additional heat brought in by the cooling device. Even at an ambitious efficiency of 10 %, we would talk about a tenfold increase in local power dissipation just to remove heat from a particular spot. It does not seem realistic to extend the range of applications into the microelectronics. The cost of integrating new materials into existing CMOS fabrication lines alone places an important barrier for introducing molecular junctions. Furthermore, depending whether the cooling device is to be placed in the device layer in the so-called front end of line (FEOL), or in the subsequent layers in the back end of line (BEOL), the material would have to stand significant temperatures during fabrication of other parts of the chip. This so-called "thermal budget" can exceed 1000 C for the front end. Integrated thermoelectrics based on organic molecules appear out of question in this context.

More important is solid-state refrigeration for device cooling. For applications near room temperature, the existing thermoelectric materials would need to be outperformed. Application include cooling of optical detectors to guarantee performance. However, it is worth looking beyond such near-room temperature applications. Today an increasing amount of electronics and sensors are operated a cryogenic temperatures. To reach these, one typically uses liquid cryogens, which implies a certain minimum size and price of installation. A solid-state solution, even with low overall efficiency would enable additional applications. As a target, consider, for example, a cooling device of 1 liter volume, operated using 1 A, that can cool a 1 $cm^3$ device to 10 K in a room temperature environment. Such a cooler may be of interest in various context. For comparison, cascades of Peltier coolers today have been shown to exceed 100 K below an ambient environment.

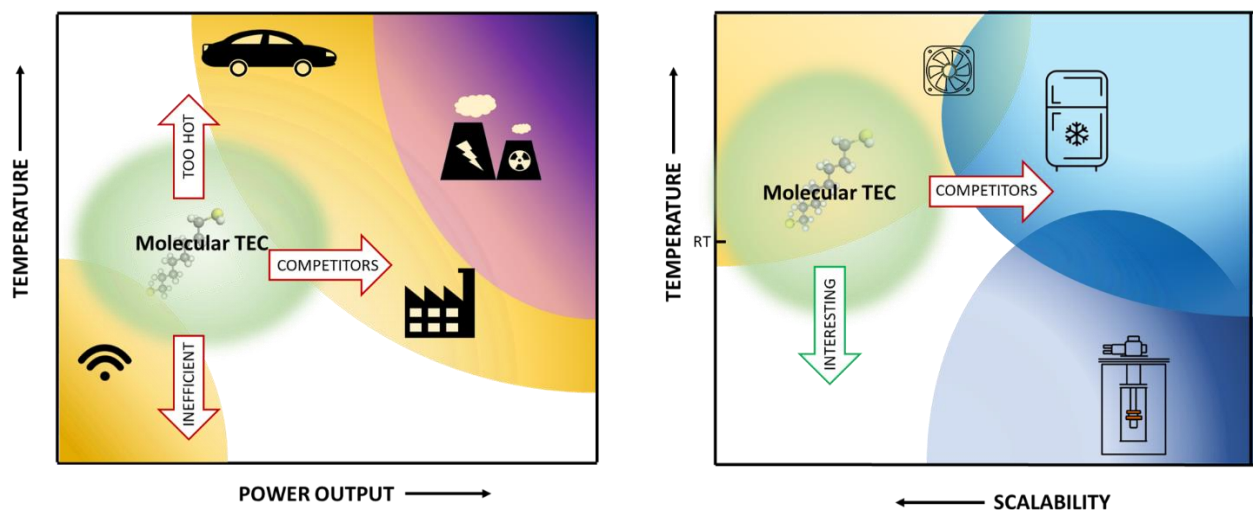

**Figure 8: Simple depiction of the use areas and challenges for cooling and harvesting applications.**

## Conclusions

Figure 8 summarizes aspects discussed here. For both energy harvesting and cooling applications, there are different applications and constraints. In terms of application space, there appears little hope for large scale (MW) and small scale (sub-mW) conversion, but moderate hope for energy harvesting at moderate power (mW to kW) and temperature (ambient to few hundred degrees C), especially in cases where maintenance and repair play a major role, as for remote sensing purpose. In the cooling space, the case for hot-spot cooling still has to be made, but there is some potential for device cooling near or below ambient temperatures. Whether molecules can compete even in these application spaces with other thermoelectric materials is yet to be shown.

This manuscript aims to indicate some important milestones that need to be reached to put molecular junctions on the map as thermoelectric energy converters. Despite the large amount of research done in the area, significant break-through still needs to come on the single-molecule level. In parallel the thin-film issue needs to be addressed. Reaching a prototype in few years would necessitate dedicated research efforts beyond the current focus on physics and chemistry. As discussed above, these will have to be motivated by serious demonstrations on the single molecule level. It is therefore likely that the milestones which are still interesting for academic research (Milestones 1 to 6) will have to yield results, before integration issues (Milestones 7 to 9) will be seriously started.

## Acknowledgements

We acknowledge funding by the European Commission H2020-FETOPEN projects 'EFINED' Grant Agreement n° 766853, and 'QuIET', Grant Agreement n° 767187.